\documentclass[smallcondensed]{svjour3}    
\usepackage{graphicx}
\usepackage{bm}
\usepackage{natbib} 
\usepackage{booktabs}
\usepackage{amssymb}
\begin{document}

\title{Can GJ 876 host four planets in resonance?}

\author{Enrico Gerlach         \\
        Nader Haghighipour }

\authorrunning{Gerlach \& Haghighipour} 

\institute{E. Gerlach \at
              Technical University of Dresden,
              Institute for Planetary Geodesy,
              Lohrmann-Observatory,
              01062 Dresden,
              Tel: +49-351-463 32050,
              Fax: +49-351-463 37019,
              \email{enrico.gerlach@tu-dresden.de}\\
           N. Haghighipour \at
           Institute for Astronomy and NASA Astrobiology Institute,
           University of Hawaii, Honolulu, HI 96822, USA,
              Tel: +1-808-956-6098,
              Fax: +1-808-956-4532,
              \email{nader@ifa.hawaii.edu}
 }
\maketitle

\begin{abstract}

Prior to the detection of its outermost Uranus-mass object, it had been suggested that 
GJ 876 could host an Earth-sized planet in a 15-day orbit. Observation, however, did
not support this idea, but instead revealed evidence for the existence of a larger body 
in a $\sim$125-day orbit, near a three-body resonance with the two giant planets of
this system. In this paper, we present a detailed analysis of the dynamics of the 
four-planet system of GJ 876, and examine the possibility of the existence of other 
planetary objects interior to its outermost body. We have developed a numerical scheme
that enables us to search the orbital parameter-space very effectively and, in a short time, identify 
regions where an object may be stable. We present details of this integration method
and discuss its application to the GJ 876 four-planet system. The results of 
our initial analysis suggested possible stable orbits at regions exterior to the orbit of 
the outermost planet and also indicated that an island of stability may exist in and around 
the 15-day orbit. However, examining the long-term stability of an object in that region by 
direct integration revealed that the 15-day orbit becomes unstable and that the system of GJ 876 is
most likely dynamically full. We present the results of our study and 
discuss their implications for the formation and final orbital architecture of this system.

\keywords{Stability \and Resonance \and Hamiltonian Systems 
\and Numerical Methods \and Planetary Systems}

\end{abstract}

\section{\label{sec:intro}Introduction}

Since the announcement of its two resonant planets by \citet{Marcy01},
the planetary system of GJ 876 has had a special place in exoplanetary science.
As the first system detected with two planets in a mean-motion resonance (MMR),
[planets GJ 876 b and c, with minimum masses of 2.27 and 0.71 Jupiter-masses and 
orbital periods of 61.11 and 30.08 days, respectively \citep{R10}, are in a 
2:1 mean-motion resonance],
GJ 876 has been the subject of extensive research and has had major contributions 
to the development of the models of planet migration and resonance capture (\citealt{Lee02},
\citealt{Kley05}). Because of the short period gravitational interactions between its two 
giant planets, the dynamics of this system and the possibility of its hosting
additional planetary bodies have been subjects of intense studies.
Soon after the announcement of its 30-day planet \citep{Marcy01},
\citet{Kinoshita01} studied the stability
of this system and showed that its 2:1 MMR presents the most stable orbital configuration
for different values of its planetary masses and orbital elements at different
epochs. \citet{Rivera01} and \citet{Laughlin01} also studied the dynamics of this system. 
These authors argued that
because of the relatively small separation between the two giant planets and the possibility 
of their close approaches, when fitting the system's radial velocities, 
the interactions between these two planets have to be taken into account.
By presenting new fits to the radial velocities of GJ 876, \citet{Rivera01} 
showed that the system will be stable for long times when the planets' orbits are co-planar
and have low inclinations with respect to the plane of the sky. 
\citet{Laughlin01} limited this inclination to a
range of $30^\circ$ to $53^\circ$. The results by \citeauthor{Rivera01} and 
\citeauthor{Laughlin01} were later confirmed by the analytical models of \citet{Ji02}
and in a thorough study of the global dynamics of GJ 876 by \citet{Gozdziewski02}.  

Being a dynamically interesting system, GJ 876 has also been the target of observation for many
years. In 2005, \citeauthor{Laughlin05} re-analyzed the new radial velocity data 
of this star and showed that when planet-planet interaction, stellar jitter, 
and instrumental uncertainties are taken into account, a 2:1 resonant co-planar
orbit with an inclination less than $20^\circ$ would present the most viable and stable
planetary configuration for the two giant planets of this system. 
These authors also showed that in addition to being in a 2:1 MMR,
the two planets are locked in a secular resonance where they librate around apsidal 
alignment with an amplitude of $34^\circ$,
and their joint line of apsides precesses at a rate of $41^\circ$ every year.
The existence of this secular resonance had also been presented in the works of 
\citet{Laughlin01}, \citet{Gozdziewski02}, 
and \citet{Lee02} where these authors analyzed the dynamics of GJ 876 
and presented a model for the migration and resonance capture of its two giant planets.

The continuous observation of GJ 876 resulted in even more fundamental discoveries. 
In 2005, \citeauthor{R05} announced that a 6.83 Earth-masses planet 
exists in a 1.94-day orbit around this star. This discovery that marked the detection
of the first super-Earth planet, prompted many astronomers to examine the possibility 
of the existence of other Earth-like bodies in this system \citep{Ji06,Ji07,Rivera07}.
In that direction, \citet{Rivera07} studied the dynamics
of fictitious planets in the system of GJ 876 and showed that in addition 
to a small region interior to the orbit of planet c where the super-Earth planet
of this system has a stable orbit, a region of stability exists beyond the orbit of the
outer giant planet (i.e., planet b) corresponding to an exterior 2:1 resonance 
with this object. In 2010, the prediction by \citeauthor{Rivera07}
materialized and a new Uranus-mass planet was discovered in a $\sim 125$-day orbit 
around GJ 876, making this system the first planetary system with three planets in a
Laplace resonance.

Prior to the detection of its fourth planet, the three-planet system of GJ 876 was studied
by \citet{Bean09} and \citet{correia2010}.
\citeauthor{Bean09} studied the architecture of the system and showed that the assumption
of co-planarity is valid, and the mutual inclination of the two giant planets are within
$1^\circ$ to $7^\circ$. In a very thorough dynamical analysis, \citet{correia2010} used 
the combined data from HARPS and Keck, and while confirming
the existence of planet d in a 1.94-day orbit, showed that the orbits of the two giant
planets of the system are co-planar with inclinations of approximately ${i_b}={48.9^\circ}$ and
${i_c}={48.1^\circ}$. These authors also suggested that an Earth-sized planet in a
15-day orbit at $\sim 0.08$ AU could have a stable orbit for a long time.
This finding was very interesting since the existence of such an additional planet 
would help to explain the anomalous high eccentricity of planet d. We would like to mention that
in a recent article, \citet{baluev2011} re-analyzed the HARPS data and suggested 
that the non-zero eccentricity of planet d may be due to the lack of proper interpretation for 
red-noise in the data.

With their new four-planet system, \citet{R10} examined the 
possibility of a stable solution for the planet proposed by \citet{correia2010}. 
However, they were unable to find a signal corresponding to the 15-day planet in their
radial velocity observations. Placing test particles in the region around 0.08 AU and integrating the
entire system numerically, these authors found that only a small fraction of particles 
remained stable for the 10 Myr duration of the integration.

Given the resonant state of the three planets of the system, and that these planets have most
probably captured each other in resonance while migrating inwards from outer regions, 
and also given the fact that GJ 876 is an M star and planet formation around M stars
favors the formation of low-mass objects, it would be natural to examine whether GJ 876
can host additional (low-mass) planets, in particular an Earth-like object in a 15-day orbit. 
This paper addresses this and several other questions regarding the dynamics of the planetary 
system of GJ 876. 

In the rest of this paper, we present our methodology and the results of our extensive
numerical analysis of the dynamics of the system. We are particularly interested in
the dynamical surrounding of the fitted orbit of planet e as given in Table 3 of \citet{R10}, 
and the possibility of additional planets in the system including
the Earth-sized planet proposed by \citet{correia2010}. We discuss our
numerical method in section \ref{sec:vareq}, and present the results in section \ref{sec:analysis}. 
Section \ref{sec:summary} summarizes 
our analysis where we also discuss their implications for the formation of this planetary system.

\section{\label{sec:vareq}Numerical tools - Stability maps from variational equations}

As mentioned above, the goal of our study is to examine the possibility of the existence of
additional bodies in the four-planet system of GJ 876. Our approach is to use stability
analysis to identify regions where an object can have a long-term stable orbit. Since there 
is no analytical solution to a general $N$-body problem with $N>2$, the orbital 
evolution of the planetary system of GJ 876 has to be computed numerically. 
As explained below, we will use a symplectic integrator for this purpose.
Due to their reliability and stability properties, especially for long-term integrations, these integrators 
have found a special place in celestial mechanics and planetary dynamics. 
We refer the reader to \citet{Hairer_etal_02} for a general overview of 
symplectic integrators and to, for instance, \citet{GDC91}, and 
\citet{ED10} for applications of these integrators to celestial mechanics.

To study the dynamical stability of an object, one has to identify (and exclude)
conditions that result in chaotic motion of the body. Several methods exist
for this purpose that discriminate between regular and chaotic motions. One can either analyze 
the orbit of a particular object [using e.g., the frequency analysis technique \citep{L93}], 
or study the evolution of deviation vectors $(\bm \delta)$ from a given orbit. 
Examples of the latter include the maximal Lyapunov Characteristic Exponent $(\gamma)$
[for a review, see e.~g. \citet{S10}], the
Fast Lyapunov Indicator technique (FLI) \citep{FLG97}, the Smaller Alignment Index (SALI)
method \citep{S01} and its generalization GALI \citep{SBA07}, and the MEGNO (Mean 
Exponential Growth of Nearby Orbits) chaos indicator \citep{CS2000}.

In this paper we concentrate on MEGNO as a fast and reliable method to study the 
dynamics of single orbits \citep{MDCG11}. For this chaos indicator, the variation 
of a quantity $Y$ and its mean value $\bar Y$ during time $t$ are given by

\begin{equation}
\label{eq:MEGNO}
 Y(t) = \frac{2}{t}\int_0^t \frac{\bm {\dot \delta}\cdot \bm \delta}
{\bm \delta \cdot \bm \delta}\,s\,\,\mathrm{d}s\qquad\mathrm{and}\qquad \bar Y(t) = 
\frac{1}{t}\int_0^t Y(s)\, \mathrm{d}s.
\end{equation}

\noindent
The quantity $\bm {\dot \delta}$ in equations (\ref{eq:MEGNO}) denotes the derivative of the deviation 
vector with respect to the independent 
variable $s$. To determine the stability of an orbit, one has to analyze the
time evolution of $\bar Y$, which is connected to the maximal Lyapunov Characteristic Exponent 
of the orbit. For stable, 
periodic orbits, $\bar Y$ approaches 0 asymptotically, while for quasi-periodic ones it will 
tend to 2. For chaotic initial conditions, $\bar Y$ will grow with time as $\gamma\,t/2$. The 
Lyapunov time, defined as $T_L=\gamma^{-1}$, can therefore be obtained from $\bar Y$ using 

\begin{equation}
\label{eq:TLyap}
 T_L = \frac{t}{2\bar Y}.
\end{equation}
\noindent
In this equation, $t$ is the total integration time.

The above-mentioned variational equations have to be integrated together with the equations of motion 
of the system. To increase the speed of these computations, we used a symplectic integrator,
and integrated the entire set of equations of the system simultaneously. 
\citet{SG2010} discuss different methods of employing
these integrators for advancing the variational equations of a Hamiltonian system.
The most efficient of these methods, the so called \emph{Tangent Map (TM)} technique, 
was applied by \citet{SG2010} and \citet{GS2010} 
mainly to low-dimensional Hamiltonian systems with 
2 and 3 degrees of freedom. \citet{GES2011} have shown that 
this method can also be applied to multi-dimensional Hamiltonian systems and is very efficient 
and superior to other commonly used numerical schemes, both in accuracy 
and speed. We note that the TM method  was first discussed in the context of celestial mechanics 
by \citet{MI1999}.

In this paper, we explore the dynamical stability of hypothetical planets in the system
of GJ 876 for a wide range of their orbital elements. To make this task computationally feasible, 
we use the SABA$_n$/SBAB$_n$ 
integrators developed by \citet{LR01}, which proved to be efficient and reliable. These symplectic 
splitting integrators have been designed specifically for integrating Hamiltonian systems of the form 
$H=A+\epsilon B$, where $A$ and $B$ are separately integrable, and $\epsilon\ll 1$, as in the case 
of hierarchical $N$-body systems. For more details on this technique and on different methods of 
splitting the Hamiltonian into 
two integrable parts, we refer the reader to e.g. \citet{DLL98}, \citet{C99}, \citet{GBB2008} 
and references therein. 

As mentioned above, we used the TM method to compute the variational equations. The formulas for 
advancing variational equations using a symplectic integrator can be found in e.g. 
\citet{MI1999} and \citet{GBB2008}. 
We used these formalisms, specifically the latter, where also
a time-discrete approximation of equations (\ref{eq:MEGNO}) is given. Let us finally note that 
symplectic methods cannot be used with a trivial automated step-size control. Therefore, they are 
usually implemented with a fixed integration step, which is denoted by $\tau$ in this paper.

\section{\label{sec:analysis}Stability analysis}

In this section, we present the results of our stability analysis for the GJ 876 planetary system. 
We used the SABA$_4$ symplectic integrator \citep{LR01} to integrate the 
equations of motion and the variational equations of the system. The latter are 
computed only for the test particles, which we define in this study to be always massless and used to determine the stability in the specific regions.
Note that all orbital elements are given with respect to the central body and integrations 
were carried out incorporating only Newtonian gravity. No relativistic and tidal effects were
included.

\subsection{\label{sec:3.1}Global stability in the inner three-planet system} of GJ 876

One of the exciting results of the dynamical analysis of GJ 876 by \citet{correia2010}
is the suggestion that a stable region exists for an Earth-sized planet at 
$\sim 0.083$ AU. This island of stability that corresponds to an orbital period of 15 days, would
allow a terrestrial planet to be in a three-body 1:2:4 resonance with the two giant
planets of the system. The interesting fact is that if this 15-day planet exists, it may be 
possible to use the interaction between this object and the innermost planet of the system 
(the 6.5 Earth-masses super-Earth that orbits the central star in a 1.94-day orbit)
as a way to account for the relatively high eccentricity of the latter body. 

Despite the predictions of \citet{correia2010}, radial velocity observations
have not been able to detect a planet in the 15-day orbit. Instead, observations by \citet{R10}
revealed that an additional planet does exist in the system, but it is Uranus-mass 
and in an orbit with a period of approximately 125 days.
This body is currently the outermost planet of GJ 876 system, and as shown
by \citet{R10}, forms a three-body Laplace resonance with the two inner giant
planets. 

The purpose of our study is to determine how the discovery of this new planet affects the stability
of the 15-day orbit. In other words, we would like to examine whether an Earth-sized planet can 
maintain a stable orbit at 0.083 AU in this new four-planet configuration? 
A few simulations by \citet{R10} suggested 
that stable motion may be possible around $\sim 0.083$ AU. However, those simulations
had been carried out only for short integration times. Our goal is to study also the long-term 
stability of orbits in this region. 

\begin{table} [hb]
  \caption{\label{tab:t1}Orbital parameters for the different models used in sections \ref{sec:3.1} and \ref{sec:3.2}.}
\centering
\footnotesize
\begin{tabular}{ccccccc}
\toprule
        &  \multicolumn{3}{c}{model I} & \multicolumn{3}{c}{model II}\\
source  & \multicolumn{3}{c}{\citet[Table 2]{correia2010}} & \multicolumn{3}{c}{\citet[Table 2]{R10}} \\
$M_\star$ $[M_\mathrm{Sun}]$ &  \multicolumn{3}{c}{0.334} & \multicolumn{3}{c}{0.32} \\
planets & b     & c     & d     &  b & c & d \\ 
\midrule
$a$ [AU]& 0.211 & 0.132 & 0.021 & 0.208 & 0.130 & 0.021 \\
$e$     & 0.029 & 0.266 & 0.139 & 0.029 & 0.255 & 0.257 \\
$i$ $[^\circ]$ & 48.93 & 48.07 & 50 & 59 & 59 & 59 \\
$\Omega$ $[^\circ]$ & 0 & -2.32 & 0 & 0 & 0 & 0 \\
$\omega$ $[^\circ]$ & 275.52 & 275.26 & 170.60 & 50.70 & 48.67 & 229.0 \\
$\lambda$ $[^\circ]$ & 35.61 & 158.62 & 29.94 & 16.10 & 343.67 & 227.0 \\
$M$ $[M_\mathrm{Jup}]$ & 2.64 & 0.83 & 0.0198 &  2.276 & 0.714 & 0.022 \\
\bottomrule
\end{tabular}
\end{table}

As we mentioned in section \ref{sec:vareq}, our approach is numerical. In order to verify that 
our numerical method produces
reliable results, we first used our code and integrated planets b, c and d together with a large 
battery of test particles
in the region interior to the orbit of the 30-day giant planet. We carried out two sets of 
simulations: one with the orbital elements for the planets as given in Table 2
of \citet{correia2010} (hereafter model I), and one with the data from \citet[Table 2]{R10} (hereafter model II). 
The integrations were done for $t=500$ years using a time-step of $\tau=0.05$. 

\begin{figure} [tb]
\centering 
\includegraphics[width=0.48\textwidth] {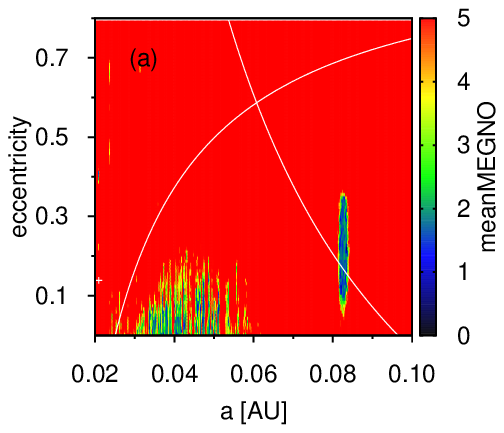} 
\hfill
\includegraphics[width=0.48\textwidth] {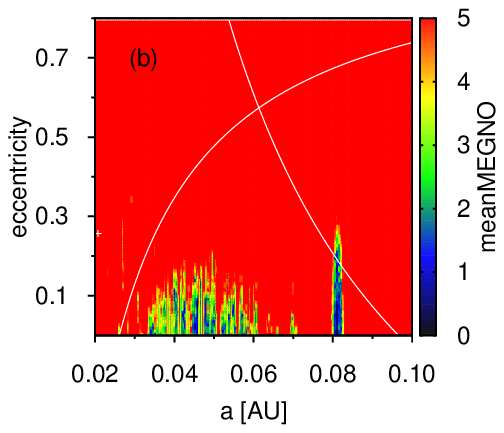} 
\caption{\label{f1}Stability graphs of test particles 
in the three-planet system of GJ 876. The color-coding represents the mean value of 
a particle's MEGNO. Red corresponds 
to chaotic motions and black/blue denote stable regions. The integrations were carried out for 500 years 
using a step-size of $\tau=0.05$ days. The initial orbital elements of the planets are given in Table~\ref{tab:t1}. Graph (a) shows the results for model I and graph (b) depicts those of model II. A grid of 600 x 100 test particles was used, for which the initial inclinations were set 
to $i=50^\circ$ in the simulations of graph (a) and they were considered to be co-planar with the 
known planets of the system, i.~e. $i=59^\circ$, in the simulations
of graph (b). All other orbital angles were initially set to zero. The white lines in both 
graphs mark the boundaries of orbit-crossing with the 30-day and 1.94-day planets of the system.}
\end{figure}

Figure \ref{f1} shows the results. In this figure, the value of the mean MEGNO associated with 
the orbit of each test particle is shown for different values of its semimajor axis 
and eccentricity. As explained before, orbits with low mean MEGNOs are considered to be stable
whereas those with high values (e.g., red color) are irregular and/or chaotic. The
left panel of Fig. \ref{f1} shows the results for the three-planet system of model I
and the right panel shows similar calculations for model II. The initial orbital inclinations of test 
particles in the left panel were set to $50^\circ$ whereas in the right panel test 
particles were considered in the same plane as the planets of the system, i.~e. $i=59^\circ$. 
All other angular elements of these objects (i.e., longitude of periastron, argument of 
ascending node, and mean-anomaly) were considered to be zero.

As shown by the two panels of Fig. \ref{f1}, an island of stability exists in the region
close to 0.08 AU. While  in both simulations, the width of this region is almost identical,
it seems that in the simulations of Fig. \ref{f1}(a), where the orbital elements of the planets 
were taken from \citet{correia2010}, the stability does not include 
orbits with eccentricities close to zero. In Fig.~\ref{f1}(b), however, we find that the 
most stable test particles in the region of 0.08 AU are the ones in circular orbits. 
The apparent instability for zero-eccentricity orbits in Fig. \ref{f1}(a) could
be attributed to the differences between the orbital elements used in the two models.
For instance, while in model I a full three
dimensional solution for the planets of the system is used, in model II, it is
assumed that all planets are on the same plane. These differences could cause slight variations in their regions of stability.

We recall that the purpose of carrying out the above-mentioned simulations and generating 
Fig. \ref{f1}(a) was to test the reliability of our numerical method. A comparison between this figure 
and Fig. 10 of \citet{correia2010} indicates that our integrator is 
in fact reliable as it has been able to correctly re-produce the main features (i.e., the 
width and location of the island of stability at 0.08 AU) of the figure published by these authors. 
The slight differences between the two results, such as the stochastic pattern 
at $a\approx 0.05$ AU, are primarily due to the intrinsic chaotic characteristic of the problem and 
the differences in the used integrating schemes. 
As shown by \citet{G09}, already existing small differences, as those resulting from the use
of different chaos indication methods, or different numerical schemes, etc., could lead to
very different stability results in areas where the phase-space is highly structured.

\subsection{\label{sec:3.2}Orbital stability in the outer three-planet system of GJ 876}

As indicated by \citet{R10}, the orbit of the recently detected
Uranus-mass planet of the system is stable for at least 400 Myr.
Test particle simulations led these authors conclude that in order for this planet to maintain
orbital stability, its orbit has to be in a Laplacian resonance with the two giant planets of
the system where its period will be in a 4:2:1 ratio with the orbital periods
of these bodies.

To verify the resonant state of the new planet (hereafter, planet e), and to quantify the 
size of its stable region, we integrated together with the massive planets of model II,
the orbits of a large number of test particles in the region exterior to the 
orbit of planet b (orbital period of $\sim$61 days) for different values of their semimajor axes ($a$) and mean
longitudes $(\lambda)$. We considered the test particles to be initially in 
circular orbits, and co-planar with the 3 planets (i.~e. $i=59^\circ$). We set 
all other orbital angular elements equal to zero. 
We integrate the system for $10^5$ orbital periods of the proposed planet e ($\approx 3400$ years) 
using a step-size of $\tau=0.1$ days.

\begin{figure} [ht]
\centering 
\includegraphics[width=\textwidth] {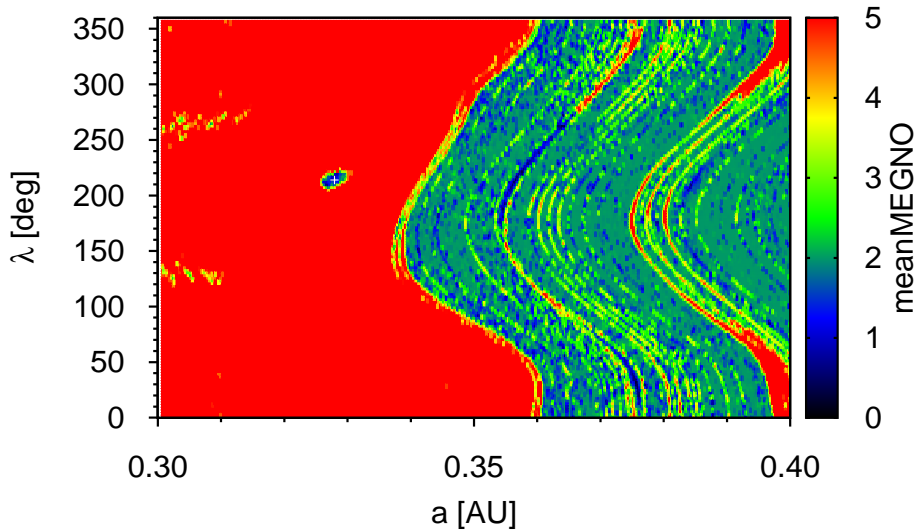} \hfill
\caption{\label{f2}Graph of the orbital stability of test particles exterior to the
orbit of planet b. Simulations were carried out for different values of the particles' 
semimajor axes and mean longitudes. The color-coding corresponds to the particles' 
mean MEGNO as described in Fig. \ref{f1}. Integrations were done for $\sim 3400$ years using 
a step-size of $\tau=0.1$ days and a grid of 400 x 150 test particles. The initial 
orbital inclinations of test particles were set to $i=59^\circ$, i.e., co-planar with the orbits 
of the three planets. However, all other orbital angular elements were initially set to zero. 
Results show a small stable region at 
$\lambda \sim {220^\circ}$ in a Laplace resonance with planets b and c,
and in the nominal position of planet $e$ (white cross) as given in Table 3 of \citet{R10}.}
\end{figure}

Figure \ref{f2} shows the results. The $(a,\lambda)$ stability map in this figure 
clearly shows the boundary between regular and chaotic orbits. Using equation (\ref{eq:TLyap}), 
one can estimate that the stable test particles in this figure have Lyapunov times of
$T_L>340$ years. An interesting feature of Fig. \ref{f2} is the small island of stability
around $a = 0.33$ AU where planet e exists. The value of $\lambda$ for this
region is $\sim 220^\circ$ which places it in a 
Laplace resonance with planets b and c. Given the small size of this area, it is clear that
this resonance protects planet e against close encounters with the other planets of the system.

Although in their analysis, \citet{correia2010} did not discuss  
the possibility of an island of stability at 0.33 AU, one can see from their Fig. 10 
that the simulations by these authors also show a small region of stability around
that area. The values of orbital eccentricity corresponding to this stable region is close
to zero which agrees with the orbital eccentricity of planet e (0.055) as reported by
\citet{R10}. This agreement between our results and the results of
\citet{correia2010} could serve as another confirmation of the validity 
of our integration method and the reliability of our results.

\subsection{\label{sec:3.3}Could an additional small planet exist in the GJ 876 four-planet system?}

As shown by our stability analysis of test particles in the three-planet 
system of GJ 876 (Fig. \ref{f1}), an island of stability seems to exist in and around 0.083 AU,
even when we use the orbital elements given by \citet{R10}. 
A planet in this region will be in or near a 2:1 MMR  with 
planet c, and as a result, will be protected against close encounters with other 
bodies. Given the orbital distance of planet e from this stable region, 
and the fact that the orbit of this object is in a mean-motion
resonance with planets b and c, it would seem natural to assume that the 
addition of planet e to the system would not alter the stability of
a planet at 0.083 AU. To examine this assumption, we carried out similar
simulations as in section \ref{sec:3.1}, but with planet e included. We used the planets' 
orbital elements as given in Table 3 of \citet{R10} (hereafter called model III, 
see Table~\ref{tab:t2}), and calculated the mean value of MEGNO for a suite of test particles
in the region between 0.02 AU and 0.1 AU. 

\begin{table} [t]
  \caption{\label{tab:t2}Orbital parameters for the model used in section~\ref{sec:3.3}.}
\centering
\footnotesize
\begin{tabular}{ccccc}
\toprule
        &  \multicolumn{4}{c}{model III} \\
source  & \multicolumn{4}{c}{\citet[Table 3]{R10}} \\
$M_\star$ $[M_\mathrm{Sun}]$ & \multicolumn{4}{c}{0.32} \\
planets &   b & c & d & e \\ 
\midrule
$a$ [AU]& 0.210 & 0.130 & 0.021 & 0.328 \\
$e$     & 0.037 & 0.256 & 0.207 & 0.045 \\
$i$ $[^\circ]$ 59 & 59 & 59 & 59 \\
$\Omega$ $[^\circ]$ 0 & 0 & 0 & 0\\
$\omega$ $[^\circ]$  43.27 & 48.74 & 234.07 & 251.36 \\
$\lambda$ $[^\circ]$ 15.88 & 343.33 & 229.38 & 214.06 \\
$M$ $[M_\mathrm{Jup}]$  2.276 & 0.714 & 0.022 & 0.046\\
\bottomrule
\end{tabular}
\end{table}

\begin{figure} [b]
\centering 
\includegraphics[width=0.48\textwidth] {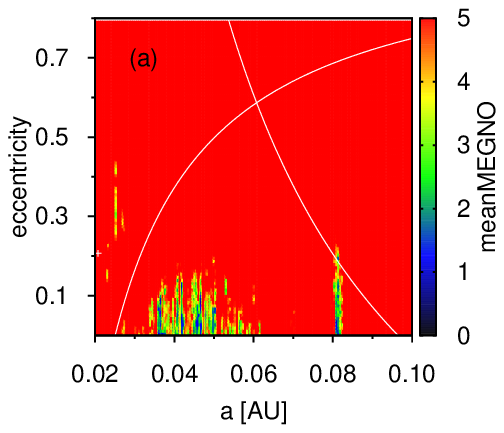} \hfill
\includegraphics[width=0.48\textwidth] {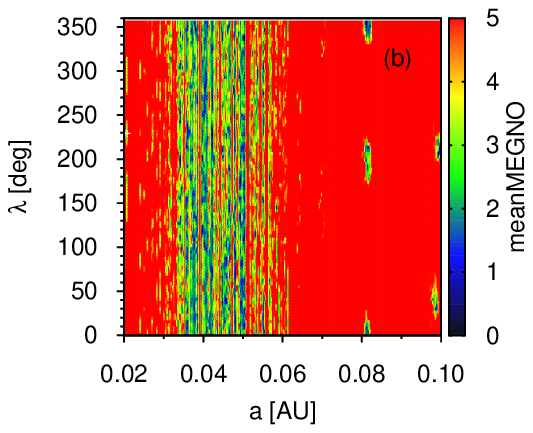}
\caption{\label{f3}Graphs of the orbital stability of test particles in the four-planet
system of GJ 876. Simulations were carried out for different values of the semimajor
axis, eccentricity, and mean longitude of test particles on a grid of 600 x 100. 
The color-coding represent 
the mean value of MEGNO as described in Fig. \ref{f1}. 
Integrations were carried out for 500 years using a step-size 
of $\tau=0.05$ days. The initial orbital elements of the planets are given in 
Table~\ref{tab:t2}. All test particles were placed 
in co-planar orbits with planets, i.~e. $i=59^\circ$. In graph (a), the initial values of other angular
variables of test particles were also set to zero. However, in graph (b), the mean anomalies
of the test particles were kept at non-zero values, but instead, their initial
eccentricities were set to zero. In (a) the white lines mark the boundary, above which 
collisions with planets b and d (crosses) are possible.}
\end{figure}

Figure \ref{f3} shows the results.
>From this figure, it can be seen that the stable island at $0.083$ AU still exists.
However, its size on the eccentricity and semimajor axes has become smaller. This figure
also shows that the region corresponding to the most stable orbits (the darkest area) 
is now exclusively for test particles in circular motion. Figure \ref{f3}(b) 
shows that the stability also crucially depends on the initial value of the mean longitude, 
which should be zero in order to place the body inside the 2:1 MMR with planet c. 

\begin{figure} [b]
\centering 
\includegraphics[width=\textwidth] {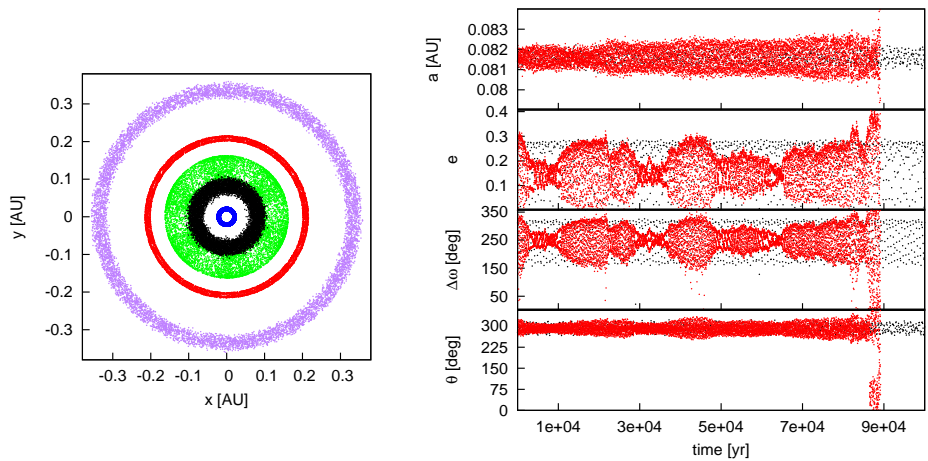}
\caption{\label{f4}Results of the long-term integration of the orbit of a test particle 
from the most stable part of the island of stability in Fig. \ref{f3}. The semimajor axis 
of the particle is initially $a=0.081420$ AU. Its other orbital elements were initially set to zero. 
The initial orbital elements of the four known planets of the system were taken from 
model III (Table~\ref{tab:t2}). The integration was carried out using a step-size of $\tau=0.05$ days. 
The graph on the left shows the orbits of the planets and that of the test particle (shown in black) 
in the $x-y$ plane. The graph on the right shows the time evolution of semimajor axis $a$, eccentricity $e$, 
difference in periastron longitudes  $\Delta \omega=\omega-\omega_c$, and resonance angle 
$\theta=-\lambda+2\lambda_c-\omega_c$ of the test particle. Shown in black color in 
the background is the orbital evolution for a test particle with the same initial 
conditions, but integrated in a system with only 3 planets as in the right panel of Fig. \ref{f1}.}
\end{figure}

Although our calculations of MEGNO suggest stability at 0.083 AU, it is important to
emphasize that this result has been obtained only from 500 years of simulations.
This time span is sufficient to identify unstable regions at reasonable CPU costs, especially 
when the grids in semimajor axis, eccentricity, and other orbital elements are very dense.  
However, it is not long enough to allow us to make conclusions
about the long-term stability of the identified stable regions. 
For the latter, integrations have to be carried out for longer times. 
In that respect, Fig. \ref{f3} can be used to 
make the search for long-term stability very efficient by allowing to focus only 
on the areas of the orbital parameter-space where an object may have a chance to be stable. 
For the four-planet system of GJ 876, this area is the 15-day orbit. We, therefore, considered a test particle 
with orbital elements corresponding to the most stable zone around 0.083 AU (the darkest spot in Fig. \ref{f3}), 
and integrated its orbit along with all other planets of the
system, for longer times. The results are shown in Fig. \ref{f4} by red color.
As can be seen in this figure, the orbit of the test particle was stable for only $9 \times 10^4$ years.
During this time, the object was trapped in a 2:1 mean-motion resonance with planet c and
its corresponding resonant angle, $\theta=-\lambda+2\lambda_{\rm c}-\omega_{\rm c}$, 
was librating around $310^\circ$. The difference between the longitude of the periastron of the 
particle $(\omega)$ and that of planet c $(\omega_{\rm c})$ at that state
was $\Delta \omega=\omega-\omega_c \sim 250^\circ$.

Figure \ref{f4} also shows that during the course of integration, strong perturbations from planet c 
caused the eccentricity of the test particle to rapidly increase. At this state, the variations
in $\Delta \omega$ (and $\theta$) developed large amplitudes until shortly before the ejection 
of the particle from the system when $\Delta \omega$ and $\theta$ began to circulate. The latter 
indicates that the orbit of the particle was no longer stabilized by the 2:1 MMR.

As mentioned before, the main purpose for carrying out these simulations
was to determine the extent to which the stability of an object in the 15-day orbit
would be affected by the newly discovered planet of the system. In that respect, and in order to be
able to make a comparison with the stability analysis of the system prior to the detection of
planet e, it is necessary to carry out similar $N$-body integrations for the 
15-day orbit test particle with only planets b, c, and d included. 
We used the orbital elements of these planets as in model II 
and integrated the orbit of the same test particle as in Fig. \ref{f3},
for 1 Myr. These results are presented as black dots in the right panel of Fig. \ref{f4}. We found
that for this configuration,
the orbit of the test particle was stable for the entire duration of the integration. 
The latter indicates
that the addition of planet e had a profound effect on the stability of this object. 
Given that the orbit of our test particle was the least chaotic one in the area around
0.083 AU, this result further suggests that the region between planets b, c, and e in the system of GJ 876
is naturally unstable. In other words, the system of GJ 876 seems to be dynamically full.

\subsection{Stability outside planet e}

\begin{figure} [ht]
\centering 
\includegraphics[width=\textwidth] {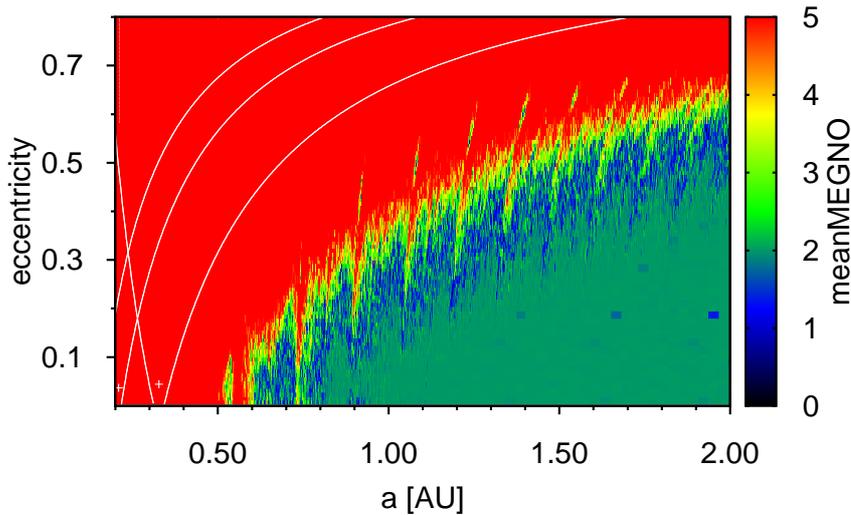} \hfill
\caption{\label{f5}Graph of the stability of test particles in the region 
exterior to planet e using a grid of 1200 x 100.
The color-coding represents the mean value of particles' MEGNO as described in Fig. \ref{f1}.
The nominal positions of the outer planets b and e are shown by white crosses. 
The white lines mark boundaries of crossing orbits with the three outer planets.
Integrations were carried out for $\sim 3400$ years using a step-size of $\tau=0.1$ days. 
The initial inclinations of test particles  were set to 
$i=59^\circ$ (i.e., co-planar with the orbits of the four planets). All the other 
orbital angles were set to zero. The initial orbital elements of the four planets 
were taken from Table~\ref{tab:t2}.}
\end{figure}

We also examined the stability of an object exterior to the orbit of planet e.
We used the orbital elements of the planets from Table~\ref{tab:t2} (model III) and carried out similar simulations
for test particles in the region of $0.2\le a \le 2.0$ AU. The results are given in Fig. \ref{f5}.
As shown there, a large stable region exists outside the orbit of planet e where the system can
host additional planets. The inner edge of this region is determined mainly by interaction
with planet e. The fine structures in this region, such as the gaps and islands of stability, are
due to various mean-motion resonances. For instance,
the large gap at 0.53 AU corresponds to a 4:1 mean motion resonance 
with planet b. Figure \ref{f5} also shows that as the semimajor axis of an object increases,
stability extends to orbits with larger eccentricities. However, the upper limit
of the eccentricity is set by the close approach of the object to planet e. 

An interesting feature of Fig. \ref{f5} is the small islands of stability that appear at high
eccentricities outside the gaps of instability that are due to mean-motion resonances. 
Whether these islands of stability are long-term stable is a topic that requires $N$-body 
integrations of an actual object in those locations.
Such a study is, however, beyond the scope of this paper.

\section{\label{sec:summary}Summary and Conclusions}

We presented the results of a detailed study of the stability of the planetary system 
of GJ 876. Our goal was to determine whether an Earth-sized planet could maintain a
stable orbit in a four-body resonance configuration with the currently known planets 
of the system. We computed the value of the chaos indicator MEGNO for a 
large number of massless test particles to map the system's
orbital parameter-space, and identified regions where the orbit of an additional
object could be stable. Results suggested an island of stability in and around
the 15-day orbit. However, direct integration of test particles showed that even orbits that were
initially very stable in that region became unstable in later times. 

We would like to emphasize that although when comparing to Jovian-type 
planets, a test particle is a good approximation for a terrestrial-class object, 
the dynamical state of a test particle may not be a true representative of the 
dynamics of an actual Earth-sized body. The mutual interactions between this
object and other bodies in the system play an important role in its
dynamical state and may result in a stable orbital configuration in systems
where test particle simulations indicate instability. 
In the case of GJ 876, however, we believe that the possibility
of the existence of such an Earth-sized planet is very small. Given the current
masses of the three giant planets of the system and their orbital configuration,
in order for a fifth planet in the 15-day orbit to be stable, the mass of this object
has to be larger than Earth-mass so that its mutual interactions with other planets
would become significant to allow its (in)stability. Given the long history of the 
observation of GJ 876  (over 12 years), such a planet is expected to have been 
discovered by now.  Results of our study suggest that the planetary system of GJ 876 
is most likely dynamically full.

It is also worth noting that from Hill radius calculations, one may 
argue that an additional,
not too massive planet with a small eccentricity can maintain stability 
in the 15-day orbit. 
Although this argument has been shown to be valid in many multiple planet 
system, our numerical integrations indicate that in the system of GJ 876, 
Hill radius analysis may not be a good approach for identifying stability. 
The latter implies that this widely used technique has
to be used with caution as it may not yield correct results at all times.

The orbital architecture of the planets around GJ 876, with a super-Earth in 
a close-in orbit and three planets in a Laplace resonance, combined with 
the fact that this system is dynamically full, raises an important and interesting
question: How did this planetary system form? Given that GJ 876 is an M star, it
is very unlikely that its outer three planets were formed in their current orbits.
In fact, as shown by \citet{Laughlin04}, the core-accretion
model of giant planet formation fails to produce Jovian-type planets around M stars.
The fact that many M stars host giant planets suggests that
these planets must have formed in outer regions of the circumstellar disk and
migrated to their current orbits. As shown by \citet{Lee02} and \citet{Kley05},
migrating planets may capture other bodies in their paths into mean-motion resonances
(mainly the 1:2 resonance) where the two (or multiple) resonant bodies continue
their migration until they either collide with the central star or
reside in a stable orbit. For a more detailed and 
comprehensive review on the formation, migration, and resonance trapping of giant 
planets around M stars, we refer the reader to \citet{Haghighipour11}. 
While migrating inward, the resonant planets excite the orbits of smaller bodies
causing many of them to be scattered out of the system. Some of these bodies may 
also collide with one another and form larger objects (e.g., super-Earths). The interactions 
between these objects and the migrating planets may result in their scattering  
to stable close-in orbits.

\begin{acknowledgements}

The authors would like to thank R. Dvorak for organizing the 8th Alexander von Humboldt 
Colloquium for Celestial Mechanics in Bad Hofgastein, Salzburg, Austria
(March 20--26, 2011) and for his kind invitation to the conference where this
project was initiated. EG would like to acknowledge support from the DFG research unit 
FOR584. NH acknowledges support from the
NASA Astrobiology Institute under Cooperative Agreement NNA04CC08A at the Institute for 
Astronomy, University of Hawaii, and from NASA/EXOB program under grant NNX09AN05G. 
The authors would also like to thank the anonymous referees for very useful 
comments and suggestions that helped to improve the clarity of the paper.

\end{acknowledgements}

\end{document}